\documentclass[twocolumn,aps,showpacs,prb,tightenlines,amsmath,amssymb]{revtex4}
\usepackage{graphicx}
\usepackage{amssymb}
\usepackage{dcolumn}
\usepackage{amsmath}
\usepackage{bm}
\usepackage{colordvi}
\usepackage{mathrsfs}
\makeatletter

\newcommand{\Rmnum}[1]{\expandafter\@slowromancap\romannumeral #1@}
\makeatother

\begin{document}

\title{Spin diffusion in $p$-type bilayer WSe$_{\bm 2}$}
\author{F. Yang}
\affiliation{Hefei National Laboratory for Physical Sciences at
Microscale, Department of Physics, and CAS Key Laboratory of Strongly-Coupled
Quantum Matter Physics, University of Science and Technology of China, Hefei,
Anhui, 230026, China}

\author{M. W. Wu}
\thanks{Author to whom correspondence should be addressed}
\email{mwwu@ustc.edu.cn.}

\affiliation{Hefei National Laboratory for Physical Sciences at
Microscale, Department of Physics, and CAS Key Laboratory of Strongly-Coupled
Quantum Matter Physics, University of Science and Technology of China, Hefei,
Anhui, 230026, China}

\date{\today}

\begin{abstract} 
We investigate the steady-state out-of-plane spin diffusion in $p$-type
bilayer WSe$_2$ in the presence of the Rashba spin-orbit coupling and
Hartree-Fock effective magnetic field. The out-of-plane components of the
Rashba spin-orbit coupling serve as the opposite Zeeman-like fields in the
two valleys. Together with the identical Hartree-Fock effective magnetic
fields, different total effective magnetic field strengths in the two valleys are
obtained. It is further revealed that due to the valley-dependent total effective
magnetic field strength, similar (different) spin-diffusion lengths in the
two valleys are observed at small (large) spin injection. Nevertheless,
it is shown that the intervalley hole-phonon
scattering can suppress the difference in the spin-diffusion lengths at large spin
injection due to the spin-conserving intervalley charge
transfers with the opposite transfer directions between spin-up
and -down holes.  Moreover, with a fixed large pure spin injection,
we predict the build-up of a 
steady-state valley polarization during the spin diffusion with the maximum
along the diffusion
direction being capable of exceeding $1~\%$. It is revealed that
the valley polarization arises from the induced quasi hot-hole Fermi
distributions with different effective hot-hole temperatures between spin-up and
-down holes during the spin diffusion, leading to the different intervalley
charge transfer rates in the opposite transfer
directions. Additionally, it is also shown that by increasing the
injected spin polarization, the hole density or the impurity density, the
larger valley polarization can be obtained.
\end{abstract}

\maketitle 

\section{Introduction}
 
In the past few years, monolayer (ML) and bilayer (BL) transition metal 
dichalcogenides (TMDs) have attracted much attention, as they provide a
promising candidate for the application in spintronics due to the two
dimensionality,\cite{2_1,2_2,2_3,inter-layer,Thick,pseudospin} gate-tunable
carrier
concentration,\cite{elecml1,electric1,electric2,Hamiltonian1,elecml2,Hamiltonian2,elecml3,elecml4,EM}
as well as the multi-valley band
structure.\cite{ML_g1,ML_g2,ML_g3,ML_g4,bandgap,ML_g5,splitting_1,ML_g6,ML_g7,splitting_2,splitting_3,ML_g8,tight-binding,electronic-structure} 
To realize the spintronic device, a great deal of efforts have been
devoted to the study on the carrier spin dynamics in this material, including the spin
relaxation\cite{s_1,s_2,s_3,s_4,s_5,s_6,s_8,s_9,s_10} and spin
diffusion.\cite{sd_1,sd_2,sd_3}  

For spin relaxation, it has been understood that the hole
spin relaxation in ML TMDs is markedly suppressed \cite{s_1,s_2,s_3} due to the
large intrinsic spin splitting.\cite{splitting_1,splitting_2,splitting_3}
As for the electron spin relaxation in
ML TMDs, the in-plane spin relaxation process has been revealed and it is
reported that the intervalley electron-phonon 
scattering makes the dominant contribution.\cite{s_4,s_5} This arises from the
intrinsic spin-orbit coupling (SOC) 
in ML TMDs, which serves as opposite out-of-plane effective magnetic fields
(EMFs) in the two valleys and hence provides the intervalley inhomogeneous
broadening\cite{KSBE,inho_1} for in-plane spins. For out-of-plane spins, the
intrinsic D'yakonov-Perel' (DP) spin relaxation process\cite{DP} in ML 
TMDs is absent due to the mirror-inversion
symmetry.\cite{ML_g1,ML_g2,ML_g3} Nevertheless, by breaking the
mirror-inversion symmetry through the flexural phonon vibrations, the
Elliot-Yafet process\cite{Yafet,Elliott} can be induced 
to cause the electron spin relaxation.\cite{s_1} In addition, with the gate-control
experimental technique on carrier
density,\cite{elecml1,electric1,electric2,Hamiltonian1,elecml2,Hamiltonian2,elecml3,elecml4} 
the external out-of-plane electric field leads to the Rashba SOC,\cite{Ra,R} and then the
extrinsic DP spin relaxation of the out-of-plane electron spins has been
predicted in ML TMDs\cite{s_2} and confirmed by the recent experiments
in ML MoS$_2$.\cite{s_6,s_9}    

Compared with ML TMDs, the intrinsic SOC in BL TMDs is absent due to the
space-inversion symmetry. This indicates that the above mentioned suppression on the
hole spin relaxation in ML TMDs is absent in BL TMDs. In the presence of
an external out-of-plane electric field $E_z$, the
experimentally realized Rashba SOC in BL TMDs can be written 
as\cite{electric2,Hamiltonian1,s_8} 
\begin{equation}
\label{Rashba}
{\bf \Omega}_{\rm R}^{\mu}({\bf k})=\big(-{\nu}k_{y},{\nu}k_{x},\mu\eta\big)E_{z},
\end{equation}
which provides a tunable out-of-plane Zeeman-like field $\mu\eta{E_z}{\bf {\hat z}}$
with opposite directions in the two valleys. Here, $\nu$ and $\eta$ are the
Rashba SOC parameters; $\mu=1~(-1)$ represents the $K$ ($K'$) valley.
For valley-independent out-of-plane spin polarization, the Zeeman-like field is
superimposed by the identical Hartree-Fock (HF) EMF $\Omega_{\rm
  HF}$\cite{HF_1,KSBE,HF_3} in each valley, leading to the larger (smaller)
total EMF $\Omega^{\mu}_{\rm T}=\mu\eta{E_z}+\Omega_{\rm 
  HF}$ in the valley possessing same (opposite) directions between
$\mu\eta{E_z}$ and $\Omega_{\rm HF}$. In our
previous work, we calculated the hole spin relaxation in BL WSe$_2$ in the presence 
of the Rashba SOC.\cite{s_8} It is pointed out that due to the presence of
the total EMF, the conventional inhomogeneous broadening in each valley is reduced by 
the magnetic field prefactor $(1+|\Omega^{\mu}_{\rm T}\tau_p|^2)^{-1}$ with
$\tau_p$ the momentum relaxation time, leading to the enhancement on
the spin relaxation time (SRT).\cite{magnetic_1,magnetic_10,magnetic_2}
Therefore, at small (large) spin polarization and hence weak (strong) HF EMF, 
identical (different) SRTs in
the two valleys are obtained.  Nevertheless, the 
intervalley hole-phonon scattering can suppress the difference in the spin
polarizations and hence the SRTs between the two valleys by inducing the
spin-conserving intervalley charge transfers with opposite
transfer directions between spin-up and -down holes. 
Therefore, via enhancing the intervalley hole-phonon scattering, the difference
in SRTs between the two valleys at large spin polarization can be markedly suppressed.
Moreover, during the spin relaxation, the quasi hot-hole Fermi
distributions with different effective hot-hole temperatures for spin-up and
-down holes are found to be induced by the spin precessions at large spin
polarization and low temperature, due to the weak hole-phonon
scattering but relatively strong hole-hole Coulomb scattering.  
With this effective hot-hole temperature difference between spin-up and -down
holes, the intervalley charge transfers said above share different rates in the
two opposite transfer directions, making the initially equal densities in the
two valleys broken (refer to Fig.~1 in Ref.~36). Hence, the valley polarization is built up. 

In contrast to the spin relaxation, the study for the spin
diffusion in ML and BL TMDs is so far rarely reported in the literature.
In ML TMDs, it has been reported that the intravalley scattering makes the dominant
contribution during the in-plane spin
diffusion whereas the intervalley one is marginal.\cite{sd_1} This is different from the  
in-plane spin relaxation,\cite{s_3} where the intervalley scattering plays an
important role. For BL TMDs, in the presence of the Rashba SOC
and HF EMF, rich physics of the out-of-plane spin diffusion can be expected. 
Specifically, the spin spatial precession frequency in each valley 
is determined by\cite{inho_2,inho_3,inho_4}
\begin{equation}
\label{inhomogeneous}
{\bm \omega}_{\bf k}^{\mu}=\frac{m}{k_x}\big({\bf \Omega}_{\rm R}^{\mu}+{\bf \Omega_{\rm
    HF}}\big)=m\big(-{\nu}E_z\frac{k_y}{k_x},{\nu}E_z,\frac{\Omega^{\mu}_{\rm T}}{k_x}\big)
\end{equation}
when the diffusion is along the ${\bf {\hat x}}$ direction.
Here, $m$ stands for the effective mass of holes. It is noted that the previous work on ultracold
$^{40}$K gas by Yu and Wu\cite{msp} shares similar 
spin spatial precession frequency [${\bm 
  \omega}({\bf k})=m(\Omega/k_x,0,\alpha)$ with $\Omega$ acting as an EMF,
$\alpha$ being the SOC strength, and the spin polarization parallel to EMF], 
except for an additional field $(m{\nu}E_z{k_y}/{k_x},0,0)$ perpendicular
to EMF in BL WSe$_2$ which provides the inhomogeneous
broadening.\cite{KSBE,inho_1,inho_2,inho_3,inho_4} Accordingly, in comparison
with the rich 
regimes of the spin diffusion in cold atoms,\cite{msp} 
different and rich spin-diffusion features in each valley
are anticipated in BL WSe$_2$. Moreover, due to the valley-dependent total EMF
strength, different 
spin-diffusion lengths in the two valleys can be obtained at the weak
intervalley scattering. Furthermore, in the presence of the spin spatial
precessions, the quasi hot-hole Fermi distributions
with different effective hot-hole temperatures between spin-up and -down holes are
expected at large spin injection and low temperature. 
Hence, similar to the induced valley polarization in the time domain as
mentioned above,\cite{s_8} one may also expect a steady-state valley polarization in
the spatial domain.     

In the present work, by the kinetic spin Bloch equation (KSBE)
approach,\cite{KSBE}  we investigate the steady-state
out-of-plane spin diffusion in $p$-type BL WSe$_2$ with 
all the relevant scatterings included. Both cases 
with and without the intervalley scattering (intervalley hole-phonon
scattering) are studied. For the case without the intervalley scattering,
the spin-diffusion processes in the two valleys are
independent, and it shown that the spin-diffusion
system in each valley can be divided into four regimes by tuning
the total EMF strength in the corresponding 
valley, similar to the rich regimes of the spin diffusion in cold atoms
mentioned above.\cite{msp} 
In each regime, the spin-diffusion length shows different dependencies on the
scattering, total EMF and SOC strengths.     
At small (large) injected spin polarization and hence the weak (strong)
HF EMF, the total EMFs, determined by the Zeeman-like fields (Zeeman-like fields
and HF EMFs), possess identical (different) strengths in the two
valleys. Therefore, similar (different) spin-diffusion lengths in the two
valleys are observed. 

When the intervalley hole-phonon scattering is included, the difference
in the spin-diffusion lengths in the two valleys is suppressed. Specifically, at
large spin injection, with the different spin diffusion lengths and
hence the different spin polarizations along the diffusion direction in the two valleys, the spin-conserving
intervalley charge transfers with opposite transfer directions
between spin-up and -down holes are triggered, which tend to suppress the
difference in the spin polarizations. The suppression is found to become stronger with the
enhancement of the intervalley hole-phonon scattering.  Moreover, with a fixed
single-side large pure spin injection, we find that a   
steady-state valley polarization along the spin-diffusion direction is built
up at low
temperature. It is further revealed that the valley polarization is induced by the
different intervalley charge transfer rates between 
spin-up and -down holes, which possess opposite transfer directions. The
difference in the intervalley charge transfer rates here arises from
the induced quasi hot-hole Fermi distributions with different effective
hot-hole temperatures between the spin-up and -down holes during the spin diffusion. 
In addition, it is found that by increasing the impurity density, the maximum
valley polarization along the diffusion direction can be markedly enhanced. This
is very different from the time domain, in which the maximum
valley polarization is always suppressed with the increase of the intravalley
scattering strength. With the physics of this unique enhancement
further revealed, it is shown that larger valley polarization can be reached by
increasing the hole density and/or injected spin polarization at large impurity
density. Particularly, at the experimental obtainable hole density and injected
spin polarization, we report
that the maximum valley polarization along the diffusion direction can
exceed $1~\%$, providing
the possibility for the experimental detection.

This paper is organized as follows. In Sec.~{\ref{model}}, we introduce our model and lay
out the KSBEs. Then in Sec.~{\ref{intravalley}}, we study the out-of-plane
spin diffusion both analytically and numerically without the intervalley hole-phonon
scattering. In Sec.~{\ref{intervalley}}, we show the influence of the
intervalley hole-phonon scattering on the out-of-plane spin diffusion. The
investigation of the induced valley polarization during the spin diffusion is
also addressed in this part. We summarize in Sec.~{\ref{summary}}.

\section{MODEL AND KSBEs}
\label{model}
In the presence of an out-of-plane electric field,
the effective Hamiltonian of the lowest two hole bands near the $K$
($K'$) point in BL WSe$_2$ is given by\cite{electric2}
\begin{equation}
\label{Hamiltonian0}
H_{\rm eff}^{\mu}=\varepsilon_{{\bf k}}+{\bf \Omega}_R^{\mu}\cdot{\bf s},
\end{equation}
where $\varepsilon_{{\bf k}}={\bf k}^2/(2m)$; ${\bf s}$ denotes the spin vector
and the Rashba SOC ${\bf \Omega}_{\rm R}^{\mu}$ is given in Eq.~({\ref{Rashba}}). 

The microscopic KSBEs, constructed to investigate the hole spin diffusion in BL
TMDs, can be written as\cite{KSBE}
\begin{equation}
  \label{KSBEs}
\dot{\rho}_{\mu{\bf k}}({\bf r},t)=\dot{\rho}_{\mu{\bf k}}({\bf r},t)|_{\rm
  coh}+\dot{\rho}_{\mu{\bf k}}({\bf r},t)|_{\rm diff}+\dot{\rho}_{\mu{\bf k}}({\bf r},t)|_{\rm scat},
\end{equation}
where $\dot{\rho}_{\mu {\bf k}}({\bf r},t)$ represent the time derivatives of the density matrices of hole
with momentum ${\bf k}$ at position ${\bf r}=(x,y)$ and time $t$, in which the
off-diagonal elements $\rho_{\mu{\bf k},\sigma-\sigma}$ describe the spin 
coherence and the diagonal ones $\rho_{\mu{\bf k},\sigma\sigma}$ represent the
hole distribution functions. 

The coherent terms,\cite{coh} describing the spin precessions of holes due to
the Rashba SOC ${\bf \Omega^{\mu}_{R}}$ and the HF EMF ${\bf \Omega_{\rm HF}}$,
are given by  
\begin{equation}
\dot{\rho}_{\mu{\bf k}}({\bf r},t)|_{\rm coh}=-i\big[{\bf \Omega}^{\mu}_{\rm
  R}\cdot{\bf s}+{\bf \Omega}_{\rm HF}\cdot{\bf s},{\rho}_{\mu{\bf k}}\big],
\end{equation}
where $[~,~]$ denotes the commutator. The HF EMF, from the Coulomb HF self-energy,\cite{KSBE,HF_1,HF_3,magnetic_2} reads    
\begin{equation}
\label{HF}
{\bf \Omega}_{\rm HF}({\bf k})={-\sum_{\bf k'}V_{{\bf k}-{\bf k'}}{\rm Tr}\big[\rho_{\mu{\bf k'}}{\bm \sigma}\big]},
\end{equation}
with $V_{{\bf k}-{\bf k'}}$ being the screened Coulomb potential. It is
noted that for valley-independent spin injection, the HF EMFs are identical in
the two valleys. 
The diffusion terms for the spin diffusion along the ${\bf {\hat x}}$
direction are written as   
\begin{equation}
\dot{\rho}_{\mu{\bf k}}({\bf r},t)|_{\rm diff}=-(k_x/m)\partial_x\rho_{\mu{\bf k}}({\bf r},t).
\end{equation}
For the scattering terms $\dot{\rho}_{\mu{\bf k}}({\bf r},t)|_{\rm scat}$,
we include all the relevant scatterings, i.e., the hole-hole
Coulomb, long-range hole-impurity, intravalley hole-in-plane-acoustic-phonon,
hole-in-plane- and hole-out-of-plane-optical-phonon   
and the intervalley hole-$K_{6}^{\rm L}$- and
hole-$K_{6}^{\rm H}$-phonon scatterings. All these scatterings are the spin
conserving ones. Here, $K_{6}^{\rm L}$ ($K_{6}^{\rm H}$) is
the phonon mode at the $K$ point corresponding to the irreducible representation
$E^{\prime\prime}_{2}$ of group $C_{3h}$ with the lower (higher) phonon
energy.\cite{s_1} The detailed expressions of the above scatterings and the corresponding
scattering matrix elements are given in our previous work.\cite{s_8} 

In the numerical calculation, the KSBEs are 
solved by taking the fixed double-side 
boundary conditions\cite{inho_2}
\begin{eqnarray}
  \label{side}
&&\rho_{\mu{\bf k}}(x=0,t)=\frac{f_{\mu{\bf k}{\uparrow}}+f_{\mu{\bf
      k}{\downarrow}}}{2}+\frac{f_{\mu{\bf k}{\uparrow}}-f_{\mu{\bf
      k}{\downarrow}}}{2}\sigma_z,~k_x>0,\nonumber\\
&&\rho_{\mu{\bf k}}(x=L,t)=f^0_{\mu{\bf k}},~~~~~k_x<0,
\end{eqnarray}
with the spin injection from the left side. 
Here, $f_{\mu{\bf k}\sigma}=\{\exp[(\epsilon_{\bf
  k}-\mu_{\mu\sigma})/(k_BT)]+1\}^{-1}$ with $\mu_{\mu\sigma}$ being the chemical
potential determined by the hole density and the injected spin polarization
$P^0_s$; $f^0_{\mu{\bf k}}$ is the Fermi distribution at 
equilibrium. For these boundary conditions, the states with $k_x>0$ at the
left edge $x=0$ are assumed to
be the source of the spin injection. The sample length $L$ is chosen to be
large enough (far larger than the spin-diffusion length) so
that the spin polarization vanished before it reaches the right edge. 
States with $k_x<0$ ($k_x>0$) in the interior ($0<x<L$) are determined from the
right (left) side of the sample with zero (fixed injected) spin polarization.
The hole densities are equal in the two valleys at $x=0$, and hence no
valley-polarization injection occurs. Moreover, with the same hole
density and hence same chemical potential in the
system, no charge diffusion occurs. All the material parameters used in our calculation are given in
Ref.~36. The Fermi energy in the calculation is chosen to be larger than
the effective Rashba SOC energy.

\section{INTRAVALLEY PROCESS}
\label{intravalley}
It is noted that in each valley, 
the spin spatial precession frequency [Eq.~(\ref{inhomogeneous})] is very
similar to that in the previous work on ultracold $^{40}$K gas by Yu and
Wu\cite{msp} except for 
an additional field $m{\nu}E_z({k_y}/{k_x},0,0)$ in BL WSe$_2$. Accordingly,
similar to the rich regimes of the spin diffusion in cold atoms, rich
intravalley spin-diffusion features are anticipated in BL WSe2.  
In our calculation, it is found that the intervalley hole-phonon
scattering is marginal at small spin injection and becomes important only at
large spin injection. This indicates that the
spin-diffusion is determined by the intravalley process at small spin injection. Therefore, 
in this section, we first investigate the steady-state out-of-plane spin
diffusion at small spin injection without the intervalley hole-phonon
scattering. The case at large spin
injection without the intervalley hole-phonon scattering is also
addressed in this section to facilitate the understanding of the complete
picture in the next section. Features of the spin-diffusions in the two valleys
in this section are independent and determined solely by the intravalley
spin-diffusion processes.  

\subsection{Analytical results}
\label{ana}
We first focus on the analytical study by simplifying the KSBEs [Eq.~(\ref{KSBEs})] with only the
hole-impurity scattering in the scattering terms. 
In the steady-state, the Fourier components of the density matrix
with respect to $\theta_k$ are given by 
\begin{eqnarray}
\label{order}
&&{k\nu{E_z}}\big(\big[s_-,\rho^{l-1}_{\mu{k}}\big]-\big[s_+,\rho^{l+1}_{\mu{k}}\big]\big)/2 
-i\Omega^{\mu}_{\rm T}\big[s_z,\rho^{l}_{\mu{k}}\big]\nonumber\\
&&\mbox{}=k/(2m){\partial_{x}}\big(\rho^{l-1}_{\mu{k}}+\rho^{l+1}_{\mu{k}}\big)+{\rho^l_{\mu{k}}}/{\tau_{k,l}},
\end{eqnarray}
with ${\tau^{-1}_{k,l}}=\frac{N_{\rm i}m}{2\pi}\int^{2\pi}_0d\theta_{\bf
  k}|V_{\bf k-k'}|^2(1-\cos{l\theta_{\bf k}})$ and $N_{\rm i}$
being impurity density.  

In the strong ($l_{\tau}{\ll}l_{\nu},l_{\Omega^\mu_{\rm T}}$) and
moderate ($l_{\Omega^\mu_{\rm T}}{\ll}l_{\tau}{\ll}l_{\nu}$) scattering regimes,
one only needs to keep the lowest two orders ($l=0,1$),\cite{keep} and obtains the
analytical solution for the spin polarization along the diffusion direction
from Eq.~(\ref{order}) (refer to the Appendix~\ref{A}). Following the previous work on
ultracold $^{40}$K gas,\cite{msp} by incorporating the additional field
$(m{\nu}E_z{k_y}/{k_x},0,0)$ in BL WSe$_2$, we define 
three characteristic lengths: the mean free path $l_{\tau}=k\tau_p/m$, the SOC length
$l_{\nu}=1/|{\nu}E_M|$ and the total EMF length
$l_{\Omega^{\mu}_{\rm T}}=k/|\Omega^{\mu}_{\rm T}m|$ in each valley, and show
that the spin-diffusion system can be divided into
four regimes: I, the large total EMF and moderate scattering regime
($l_{\tau}{\ll}l_{\Omega^{\mu}_{\rm T}}{\ll}l_{\nu}$); 
II, the large total EMF and strong scattering regime
($l_{\Omega^{\mu}_{\rm T}}{\ll}l_{\tau}{\ll}l_{\nu}$); III, the crossover regime ($l_{\tau}{\ll}l_{\nu}{\ll}l_{\Omega^{\mu}_{\rm
    T}}{\ll}2l^2_{\nu}/l_{\tau}$); IV, the small total EMF regime
($l_{\tau}{\ll}l_{\nu}{\ll}2l^2_{\nu}/l_{\tau}{\ll}l_{\Omega^{\mu}_{\rm
    T}}$). In different regimes, the spin polarizations exhibit different decay
behaviors and the corresponding decay lengths show different dependencies on the
scattering, SOC and total EMF strengths. The specific spin-polarization behaviors
in each regime are summarized in Table~\ref{divi}.  

It is noted that from
Eq.~(\ref{inhomogeneous}), the 
direction of the inhomogeneous broadening ${\bf {\hat z'}}$, given by  
\begin{equation}
\label{direction}
{\bf {\hat z'}}=\frac{1}{\sqrt{1+|\Omega^{\mu}_{\rm T}/(\nu{k}E_z)|^2}}{\bm
  {\hat \theta_{k}}}+\frac{\Omega^{\mu}_{\rm T}/(\nu{k}E_z)}{\sqrt{1+|\Omega^{\mu}_{\rm T}/(\nu{k}E_z)|^2}}{\bf
  {\hat z}},
\end{equation}
is nearly along the ${\bf {\hat z}}$ direction in the large total EMF regimes
(regimes I and II with $|\Omega^{\mu}_{\rm
  T}|/|\nu{k}E_z|{\gg}1$). Therefore, the out-of-plane spins cannot precess
around 
the inhomogeneous broadening effectively. In this situation, the spin polarization
decays without any oscillation, and through the modified drift-diffusion model
[$l_s=\sqrt{D\tau_s}$ with the diffusion coefficient\cite{sdd1,sdd2} $D=v_F^2\tau_p/3$ ($v_F$
represents the Fermi velocity) and SRT $\tau_s=(1+|\Omega^{\mu}_{\rm
  T}\tau_p|^2)/(|{\nu}kE_z|^2\tau_p)$] proposed by Yu and Wu in cold
atoms,\cite{msp} the spin diffusion in our work can be understood well.

As for the crossover regime (regime III) and the
small total EMF regime (regime IV), with $|\Omega^{\mu}_{\rm T}|/|{\nu}kE_z|{\ll}1$, the 
direction of the inhomogeneous broadening ${\bf {\hat z'}}$
[Eq.~(\ref{direction})] deviates from that of the out-of-plane
spin polarization, and hence the efficient spin precessions are induced. 
It has been pointed out by Yu and Wu\cite{msp}
that the modified drift-diffusion model fails to
explain the spin-diffusion in this situation.
In the present work, we suggest a reasonable
picture based on the previous works in semiconductor\cite{Si} and graphene\cite{graphene} to
facilitate the understanding of the spin diffusion in regimes III and
IV. Specifically, as seen from
Eq.~(\ref{inhomogeneous}), there are two channels
for the out-of-plane spin diffusion: (i) through the inhomogeneous broadening
provided by the conventional Rashba SOC, i.e., the additional field
$m{\nu}E_z({k_y}/{k_x},0,0)$;\cite{graphene} (ii) by rotating out-of-plane spins into the
in-plane direction via spin spatial precessions and then through the
inhomogeneous broadening for in-plane spins provided by the total
EMF.\cite{Si} In the crossover regime (regime III), both channel (i) and (ii) are
important. Nevertheless, the presence of the total EMF
suppresses the out-of-plane spin precessions induced by the conventional Rashba
SOC, and the suppression should decrease with the increase of
$|{\nu{E_z}k}/{{\Omega^{\mu}_T}}|^2=l^2_{\Omega^{\mu}_T}/l^2_{\nu}$
according to Eq.~(\ref{direction}). Therefore,
the spin polarization through channel (i)  
shows single-exponential decay with the decay length 
$l^s_s\approx{l_{\nu}}(1-2l^2_{\Omega^{\mu}_T}l^2_{\tau}/l^4_{\nu})/\sqrt{2}$.
In addition, the spin polarization 
through channel (ii) shows the oscillatory decay with the decay length
$l^o_{s}\approx\sqrt{l_{\tau}l_{\Omega^{\mu}_{\rm 
      T}}}$.\cite{Si} Consequently, the spin polarization in regime III is
approximated by one oscillatory decay together with one single-exponential
decay. With further decreasing the total EMF, the system enters
the small total EMF regime (regime IV). In this regime, due to the weak total
EMF, the inhomogeneous broadening in channel (ii) becomes inefficient, while the
suppression from the total EMF on channel (i) also becomes weak. Consequently,
the spin polarization, only determined by channel (i) without any suppression,
shows the oscillatory decay with the decay length
$l_s=l_{\nu}/(2\sqrt{2\sqrt{2}-1})$, same as the work in graphene.\cite{graphene}

\begin{widetext}
\begin{center}
\begin{table}[htb]
\caption{Behaviors of the steady-state out-of-plane spin polarization along the
  diffusion direction and the corresponding spin-diffusion lengths in each
  regime.  $l_c={2}l^2_{\nu}/l_{\tau}$. }  
\label{divi} 
\begin{tabular}{l l l l}
    \hline
    \hline
Regime&\quad~Condition&\quad\quad~Behavior&\quad~Decay length $l_s$\\
    \hline
    I:Large total EMF and moderate scattering
    regime&\quad$l_{\Omega^{\mu}_{\rm
        T}}{\ll}l_{\tau}{\ll}l_{\nu}$&\quad\quad~single-exponential
    decay&\quad$l_{\tau}l_{\nu}/(\sqrt{6}l_{\Omega^{\mu}_{\rm T}})$\\
II:Large total EMF and strong scattering
regime&\quad$l_{\tau}{\ll}l_{\Omega^{\mu}_{\rm
    T}}{\ll}l_{\nu}$&\quad\quad~single-exponential decay&\quad$l_{\nu}(1+l^2_{\tau}/l^2_{\Omega^{\mu}_{\rm
    T}})/{\sqrt{2}}$\\
III: Crossover
regime&\quad$l_{\tau}{\ll}l_{\nu}{\ll}l_{\Omega^{\mu}_{\rm
    T}}{\ll}l_c$&\quad\quad~single-exponential decay&\quad${l_{\nu}}(1-2l^2_{\Omega^{\mu}_{\rm
    T}}/l^2_c)/{\sqrt{2}}$\\
\quad&\quad&\quad\quad~oscillatory
decay&\quad$\sqrt{l_{\tau}l_{\Omega^{\mu}_{\rm T}}}$\\
IV:Small total EMF
regime&\quad$l_{\tau}{\ll}l_{\nu}{\ll}l_c{\ll}l_{\Omega^{\mu}_{\rm T}}$&\quad\quad~oscillatory decay&\quad$l_{\nu}/(2\sqrt{2\sqrt{2}-1})$\\
    \hline
    \hline
\end{tabular}\\
\end{table}
\end{center}
\end{widetext}

\subsection{Numerical results}
We next discuss the spin diffusion without the intervalley
scatterings by numerically solving the KSBEs at small and large spin
injections. To compare with the analytical results revealed in 
Sec.~\ref{ana}, both cases with only the long-range hole-impurity scattering and
with all the intravalley scatterings are studied.   
  
\subsubsection{Scattering strength dependence}

\begin{figure}[htb]
  {\includegraphics[width=9cm]{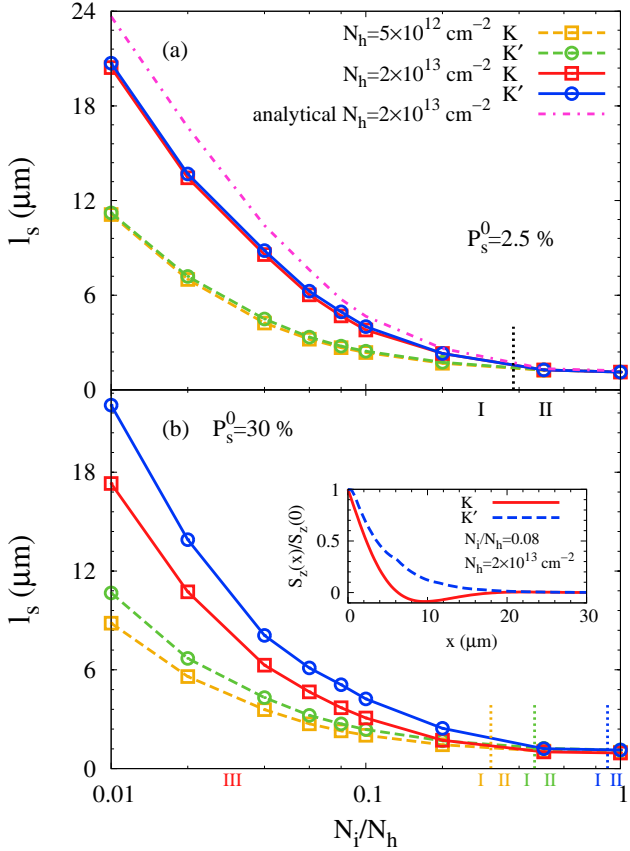}}
\caption{(Color online) Scattering dependence of the spin diffusion
  length with only the long-range hole-impurity
  scattering at
  different hole densities when (a) $P^0_s=2.5~\%$ and (b) $P^0_s=30~\%$. 
  Squares (Circles): in the $K$ ($K'$) valley. The chain curve in the figure is
  calculated from the analytical result [obtained from Eq.~(A6) by setting $\Omega^{\mu}_{\rm
    T}=\mu\eta{E_z}$]. The dotted lines on the frames indicate
  the boundaries between regimes I and II. Particularly, the boundaries between
  regimes I and II in (a)
  for different curves are located at the same position. The roman numbers with the color
  at the bottom frame indicate the regimes of the corresponding systems denoted by
  the same color. The
  inset in (b) shows the spin polarizations along the diffusion direction in the $K$ 
  (solid curve) and $K'$ (dashed curve) valleys.  $E_z=0.02~$V/{\r A}. }    
\label{figyw1}
\end{figure}

In this part, we address the scattering strength dependence of the intravalley
spin-diffusion process. We first focus on the case with
only the long-range hole-impurity scattering. The spin diffusion lengths as
function of impurity density at different hole densities are plotted in
Figs.~\ref{figyw1}(a) and (b) with small ($P^0_s=2.5~\%$) and large
($P^0_s=30~\%$) injected
spin polarizations, respectively.  At small injected spin polarization and hence the weak HF
EMF, the total EMFs, 
determined by the large Zeeman-like fields, have identical strengths in the
two valleys. Consequently, from Table~\ref{divi}, the systems in the $K$ and $K'$ valley both sit
in the large total EMF regimes and same spin-diffusion lengths
in the two valleys are obtained. When 
$N_{\rm i}/{N_h}<0.2$, the system lies in the moderate
scattering regime (regime I), and the spin-diffusion
length $l_s\propto\tau_p|\Omega^{\mu}_{\rm T}|$. It is noted that the hole-impurity scattering
strength $1/\tau^{\rm i}_p\propto{N_{\rm i}/N_h}$.\cite{Fer} Therefore, 
the increase of $N_i/N_h$ leads to the decrease of the spin diffusion
length when $N_{\rm i}/{N_h}<0.2$, as shown in Fig.~\ref{figyw1}(a). By further
increasing the scattering strength to the strong scattering regime (regime
II), the spin-diffusion length $l_s{\approx}l_{\nu}/\sqrt{2}$ becomes
scattering-independent when $N_{\rm i}/N_h>0.2$. Moreover, some marginal
difference in the spin diffusion lengths between the two valleys is observed
in the moderate scattering regime. This is due to the weak HF EMFs at small spin injection,
which lead to the small difference in the total EMFs. Additionally, identical spin
diffusion lengths in the two valleys are obtained from the analytical results
[obtained from Eq.~(A6) by setting $\Omega^{\mu}_{\rm
    T}=\mu\eta{E_z}$], as shown by the chain curve in
Fig.~\ref{figyw1}(a). It is found that the analytical results agree with the
numerical ones (solid curves) fairly well in the strong 
scattering regime and are very close to the numerical ones in the
moderate scattering regime.   

At large injected spin polarization and hence the strong HF EMF, the total EMFs
have different strengths in the two valleys, leading to different spin-diffusion
lengths according to Table~\ref{divi}.
Specifically, in our calculation, the HF EMF and Zeeman-like field have 
the opposite (same) directions in the $K$ ($K'$) valley, and hence
the total EMF has a larger strength in the $K'$ valley. At
$N_h=5\times10^{12}~{\rm cm}^{-2}$ (dashed curves), the HF EMFs ($\Omega_{\rm 
  HF}\approx1.86~$meV) are relatively
smaller than the Zeeman-like fields ($\eta{E_z}=10.6~$meV), and hence systems in
the $K$ and $K'$ 
valleys sit in the large total EMF regimes (regimes I and II). Consequently,
when $N_{\rm i}/{N_h}<0.2$ (regime I with $l_s\propto\tau_p|\Omega^{\mu}_{\rm T}|$), the
spin-diffusion length in the $K'$ valley (curves with circles) is larger than that in
the $K$ one (curves with squares), as shown in
Fig.~\ref{figyw1}(b). At $N_h=2\times10^{13}~{\rm cm}^{-2}$, the HF 
EMF ($\Omega_{\rm HF}=7.95~$meV) is relatively strong. In this
situation, the system in the $K$ ($K'$) valley lies in the crossover regime
(large total EMF regimes). As shown in the inset of
Fig.~\ref{figyw1}(b), the spin polarization in the $K$ ($K'$) valley [solid
  curves (chain curves)] shows the oscillatory (single-exponential) decay
along the diffusion direction, consistent with the analytical results. Moreover,
according to Table~\ref{divi}, it is found that the spin-diffusion length
in the $K'$ valley is also larger than that in the $K$ one, and the 
spin-diffusion length in the $K$ ($K'$) valley decreases with the increase of the
scattering strength.

\begin{figure}[htb]
  {\includegraphics[width=9cm]{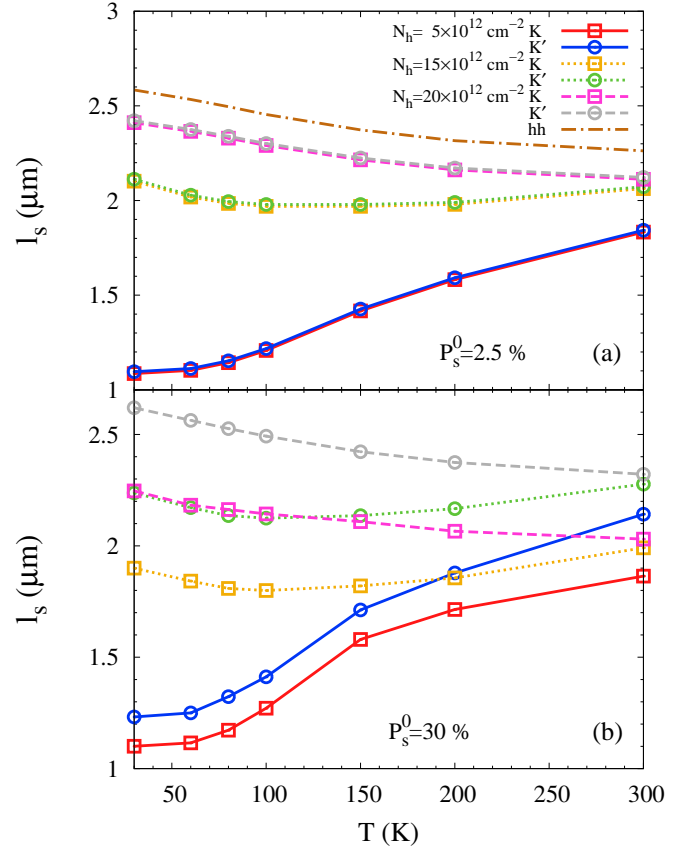}}
\caption{(Color online) The spin diffusion length versus temperature $T$ at
  different hole densities when (a) $P^0_s=2.5~\%$ and (b)
  $P^0_s=30~\%$. Squares (Circles): in the $K$ ($K'$) valley with all the
  intravalley   
  scatterings included. Chain curve: with only the hole-hole Coulomb
  scattering included at $N_h=2\times10^{13}~{\rm cm}^{-2}$. $E_z=0.02~$V/{\r
    A}.}      
\label{figyw2}
\end{figure}

We next take all the relevant intravalley scatterings (hole-hole Coulomb,
long-range hole-impurity and intravalley hole-phonon scatterings) into account. The 
impurity density is taken to be $N_{\rm i}=0.02N_{h}$ according to
Ref.~36.  The spin diffusion lengths as function of 
temperature at different hole densities are  
plotted in Figs.~\ref{figyw2}(a) and (b) with small ($P^0_s=2.5~\%$) and large
($P^0_s=30~\%$) injected spin polarizations, respectively. As seen from the figure,
different (nearly identical) spin diffusion 
lengths in the two valleys are obtained at large (small) injected spin polarization. 
In addition, it is found that in each valley, 
with the increase of the temperature, the spin diffusion lengths at small and
large spin polarizations both increase at low hole density
$N_h=5\times10^{12}~{\rm cm}^{-2}$ (solid curves) but decrease at high hole
density $N_h=2\times10^{13}~{\rm cm}^{-2}$ (dashed curves). This is due to the dominant
hole-hole Coulomb scattering in BL WSe$_2$. Specifically, as shown in 
Fig.~\ref{figyw2}(a), the spin-diffusion
length at $N_h=2\times10^{13}~{\rm cm}^{-2}$ with all the intravalley scattering
(dashed curves) is close to that with 
only the hole-hole Coulomb scattering (chain curve). This indicates that the
hole-hole Coulomb scattering makes dominant contribution in the spin
diffusion. Moreover, from the results with only the hole-impurity
scattering (Fig.~\ref{figyw1}), it has been demonstrated that in each
valley, with the decrease of $\tau_p$, the spin-diffusion
lengths at small and large spin injection both decrease monotonically before the
scattering becomes very strong, and then saturate around $l_s\approx{l_{\nu}}/\sqrt{2}$
($l_{\nu}/\sqrt{2}\approx0.82~\mu$m). When all the intravally 
scatterings are included, as shown in Figs.~\ref{figyw2}(a) and (b), the
spin-diffusion length in each valley is larger than $l_{\nu}/\sqrt{2}$. Consequently,
with the dominant hole-hole Coulomb scattering strength $1/\tau^{\rm hh}_{p}{\propto}\ln(T_{F}/T)T^{2}/T_{F}$
($1/\tau^{\rm hh}_{p}{\propto}1/T$) at
$T{\ll}({\gg})T_{F}$,\cite{pw_3,pw_3.0} the spin diffusion length decreases
(increases) with the increase of temperature in the degenerate
(nondegenerate) limit for high (low) hole density. Therefore, as shown in
Figs.~\ref{figyw2}(a) and (b), 
at hole density $N_h=1.5\times10^{13}~{\rm cm}^{-2}$ (dot curves) 
with $T_F\approx410~$K, a valley shows up at the
crossover from the degenerate to nondegenerate limits  ($T_c\approx{T_F}/4$ in
$p$-type BL WSe$_2$\cite{s_8}) in the temperature dependence of the spin-diffusion
length.

\subsubsection{Total EMF dependence}

\begin{figure}[htb]
  {\includegraphics[width=9cm]{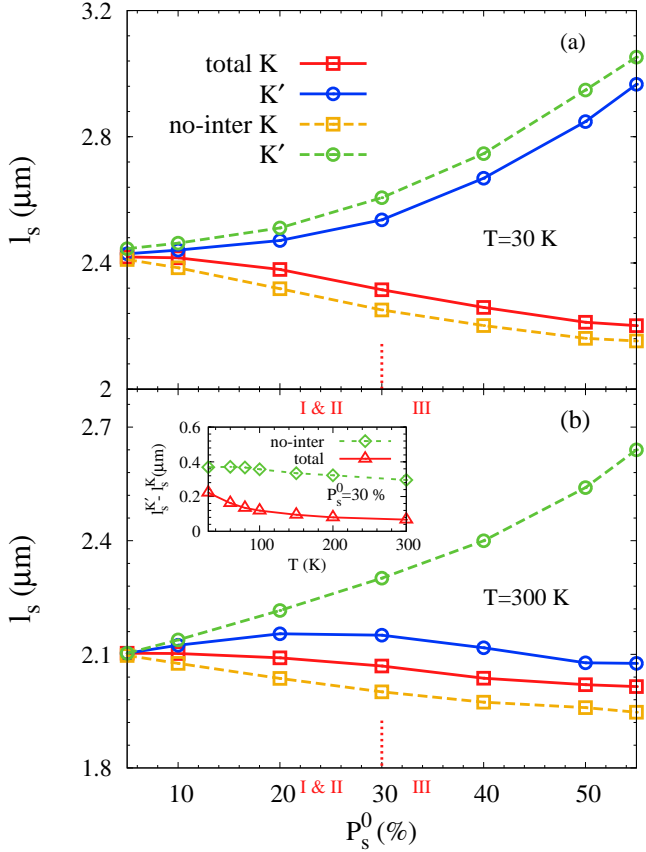}}
\caption{(Color online) The spin diffusion length as function of injected spin
  polarization $P^0_s$
  when (a) $T=30~$K and (b) $T=300~$K. Squares (Circles): in the $K$ ($K'$) valley.
  Solid (Dashed) curves: with (without) the intervalley hole-phonon
  scattering. The dotted lines on the frames indicate the boundaries
      between the large total EMF regimes (regimes I and II) and the crossover
      regime (regime III) in the $K'$ valley, and the roman numbers
      at the bottom frames indicate the corresponding regimes. It is noted that
      for the large total EMF regimes (regimes I and II) in the $K'$ valley
      (when $P^0_s<30~\%$), the boundary
      between regimes I and II ($|\Omega^{K'}_{\rm T}\tau_p|\approx1$) is hard to
      be determined due to the above mentioned dominant hole-hole Coulomb
      scattering in $\tau_p$. The system in the $K$ valley always sits in regime
      I. The inset in (b) shows the  
  difference in the spin diffusion lengths $l^{K'}_s-l^{K}_s$
  between the two valleys versus temperature $T$. Triangles (Diamonds): with (without) the
  intervalley hole-phonon scattering. $N_h=2\times10^{13}~$cm$^{-2}$ and $E_z=0.02~$V/{\r A}.}   
\label{figyw3}
\end{figure}

Next we turn to study the total EMF dependence of the spin diffusion length with
all the intravalley scatterings included.  The spin diffusion lengths as function
of the injected spin polarization without the intervalley scattering 
are plotted by dashed curves in Figs.~\ref{figyw3}(a) and (b) at low
($T=30~$K) and high ($T=300~$K) temperatures, respectively. 
At small injected spin polarization ($P^0_s<30~\%$), total EMF is determined by the
Zeeman-like field, and systems in the $K$ and $K'$ valley both sit in the large
total EMF and moderate scattering regime (regime I with
$l_s\propto|\Omega^{\mu}_{\rm T}|$). It has mentioned above that the
HF EMF and Zeeman-like field possess opposite (same) directions in the
$K$ ($K'$) valley in our calculation. Consequently,  
with the increase of the injected spin polarization and hence the HF EMF strength, the
total EMF strength in the $K$ ($K'$) valley becomes weaker (stronger),
leading to the decrease (increase) of the spin-diffusion length [dashed curve
with squares (circles)] when $P^0_s<30~\%$, as shown in Figs.~\ref{figyw3}(a) and
(b).   

Moreover, by further increasing the injected spin polarization, the system in the $K$
valley enters the crossover regime (regime III) when $P^0_s>30~\%$ and that in the
$K'$ one still sits in regime I, as mentioned above. Consequently, the spin-diffusion length in
the $K'$ valley (dashed curve with circles) still increases rapidly with the injected 
spin polarization. However, it is noted that with the increase
of the injected spin polarization, i.e., the decrease of the total EMF strength
in the $K$ valley, the spin diffusion length in this valley (dashed curve with squares)
decreases. This is hard to understand directly from Table~\ref{divi}, where it
is shown that both single-exponential decay and oscillatory decay of the
spin polarization can happen in regime III. As mentioned above, the decrease of
the total EMF enhances the single-exponential-decay channel [channel (i)] and suppresses the
oscillatory-decay channel [channel (ii)]. From our calculation,
with all the intravalley scatterings, it is found that channel (i) is more
important when $P^0_s>30~\%$, leading to the decrease of the spin-diffusion length
in the $K$ valley with decreasing the total EMF strength.

\begin{figure}[htb]
  {\includegraphics[width=9cm]{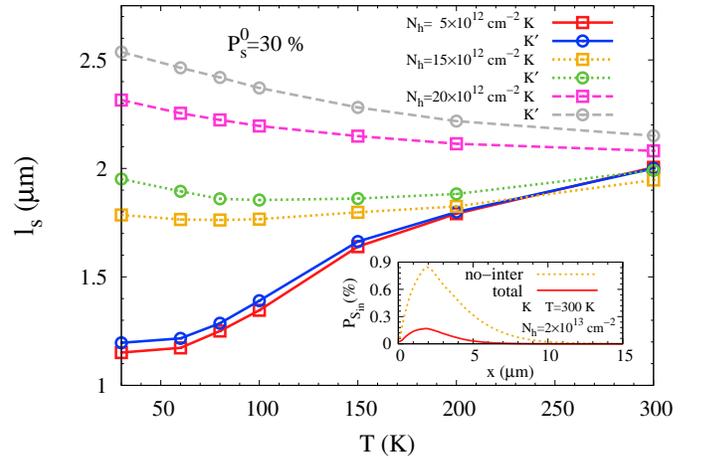}}
\caption{(Color online) The temperature dependence of the spin diffusion length
  with all the relevant scatterings included at different hole densities when
  $P^0_s=30~\%$. Squares (Circles): in the $K$ ($K'$) valley.  The
  inset shows the induced in-plane spin polarization in the $K$ valley during the
  out-of-plane spin diffusion with
  (solid curve) and without (dotted curve) the intervalley hole-phonon
  scattering. $E_z=0.02~$V/{\r A}.}     
\label{figyw4}
\end{figure}

\section{ROLE OF THE INTERVALLEY SCATTERING}
\label{intervalley}

As mentioned above, it is found that the intervalley hole-phonon
scattering is marginal at small spin injection and becomes important only
at large spin injection. Therefore, we next investigate the role
of the intervalley hole-phonon scattering on the out-of-plane
spin diffusion at large spin injection. Two aspects of the
influence are studied. 
 
On one hand, at large spin injection, with the smaller
spin-diffusion length in the $K$ valley in our calculation, the faster
decay of the spin polarization along the diffusion direction,
makes the density of spin-down (-up) holes larger (smaller) in this valley than
that in the $K'$ one, triggering the spin-conserving intervalley charge transfer of
spin-down (-up) holes from the $K$ ($K'$) valley to the $K'$ ($K$) one through
the intervalley hole-phonon scattering. Consequently, the
difference in the spin polarizations and hence the difference in the
spin-diffusion lengths between the two valleys is suppressed. 

On the other hand, it has been pointed out in the previous works\cite{s_4,s_5}
that for the in-plane 
spin relaxation (in the time domain) in ML MoS$_2$, the valley-dependent EMF
provides the intervalley inhomogeneous broadening for in-plane spins, and opens
an intervalley in-plane spin-relaxation channel in the presence of the
intervalley scattering. Similar to the time domain, the total EMF
in the spatial domain [Eq.~(\ref{inhomogeneous})] is also valley-dependent in BL
WSe$_2$, leading to the intervalley in-plane spin-decay channel during the spin
diffusion when the intervalley scattering is included.  For the out-of-plane
spin diffusion, the system in the $K$ valley at large spin injection (with
small total EMF) sits in the crossover regime, and hence the out-of-plane
spins in this valley can precess efficiently into the in-plane direction,
activating the intervalley spin-decay channel revealed above.

Finally, in the presence of the intervalley hole-phonon scattering, it is interesting
to find that a steady-state valley polarization is built up during the spin
diffusion at large spin injection and low 
temperature. We systematically investigate this interplay of the
spin polarization with the valley polarization in the spatial domain, and
find that the valley polarization arises from the quasi hot-hole Fermi
distributions with different effective hot-hole temperatures
between spin-up and -down holes, which are induced during the spin diffusion,
similar to the induced valley polarization during the spin relaxation (in the time
domain).\cite{s_8}    

\subsection{Spin diffusion}
\label{spin}
In this part, we analyze the out-of-plane spin-diffusion with both the intra-
and intervalley
scatterings. The spin diffusion lengths as function 
of the injected spin polarization with all the relevant scatterings
included are plotted by solid curves in Figs.~\ref{figyw3}(a) and (b) at low
($T=30~$K) and high ($T=300~$K) temperatures, respectively. 
With the intervalley hole-phonon scattering, the spin-conserving intervalley
charge transfer is switched on, leading to the difference in the
spin-diffusion lengths in the two valleys suppressed, as mentioned above. At low
(high) temperature $T=30~$K ($T=300~$K), as seen from
Fig.~\ref{figyw3}(a) [Fig.~\ref{figyw3}(b)], the
suppression is weak (strong) due to the weak (strong) intervalley hole-phonon scattering, and
hence different (similar) spin-diffusion lengths can be obtained. 
The differences in the
spin-diffusion lengths between the two valleys versus temperature are plotted in
the inset of Fig.~\ref{figyw3}(b) with (solid curve with
triangles) and without (dashed curve with diamonds) the intervalley hole-phonon
scattering. As seen from the inset, the suppression on
the difference in the spin diffusion lengths in the two valleys
becomes stronger with the enhancement of the intervalley hole-phonon scattering
by increasing temperature. 

In Fig.~\ref{figyw4}, we further plot the spin-diffusion length versus temperature at
large spin injection with all the relevant scatterings included. Comparing
the results with (Fig.~\ref{figyw4}) and without 
[Fig.~\ref{figyw2}(b)] the intervalley hole-phonon scattering, we find that 
the leading role of the intervalley hole-phonon scattering on the out-of-plane
spin diffusion is to suppress the difference in the spin-diffusion lengths between
the two valleys. Moreover, our study shows that the intervalley
spin-decay channel mentioned above is always
inefficient during the out-of-plane spin diffusion. This is because that the
intervalley charge transfer tends to suppress the difference in the spin
polarizations in the two valleys. With this
suppression, the efficient spin precessions in the $K$ valley due to the small
total EMF, can be effectively suppressed by the large total EMF in the $K'$ valley.
The in-plane spin polarization in the $K$ valley during the 
out-of-plane spin diffusion, which is induced by the spin spatial precessions, is
plotted in the inset of Fig.~\ref{figyw4}(b).  
As seen from the inset, in contrast to the result without the intervalley
hole-phonon scattering (dotted curve), the induced in-plane spin
polarization in the $K$ valley is markedly suppressed when the intervalley
hole-phonon scattering is switched on (solid curve), leading to the intervalley
spin-decay channel inefficient.  

\subsection{Valley polarization}

\begin{figure}[htb]
  {\includegraphics[width=9cm]{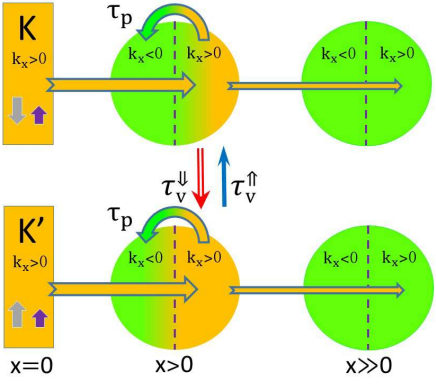}}
\caption{(Color online) Schematic of the spin-diffusion processes in the
      two 
  valleys and valley polarization process. In the figure, the purple (gray) filled
arrows, which have the same (opposite) directions in the two valleys, stand for the
HF (Zeeman-like) EMFs; the brown (green) color denotes the states with
$P_s\ne0$ ($P_s=0$). On one hand, this schematic shows that due to the
smaller total EMF and hence the smaller spin-diffusion length in the $K$ valley,
the spin polarization in this valley is smaller than that in the $K'$ one,
inducing the intervalley charge transfers with opposite transfer 
directions between the spin-up [from the $K'$ valley to the $K$ one (blue single arrow)] and -down
holes [from the $K$ valley to the $K'$ one (red double arrow)]. The intervalley
charge transfer rate $1/\tau^{\Downarrow}_{\rm v}$ for spin-down holes  is faster
than that $1/\tau^{\Uparrow}_{\rm v}$ for spin-up ones due to the larger effective
hot-hole temperature for spin-down holes (refer to Fig.~1 in Ref.~36). On the other hand, this
schematic 
exhibits that in the region away from the boundary $(x>0)$,
$P^{\mu}_{s,k_x>0}$ is induced due to the spin injection from the boundary $(x=0)$
through the $k_x>0$ states and $P^{\mu}_{s,k_x<0}$ is induced through the
scattering from the spin polarized $k_x>0$ states at the same position $x$.
}     
\label{figyw5}
\end{figure}

As mentioned in the introduction, with the different
spin-diffusion lengths in the two valleys at large spin injection,  a
steady-state valley polarization is expected to build up during the spin diffusion at low temperature, similar to the
valley polarization in the time domain.\cite{s_8} Specifically, it has
been mentioned above that the intervalley charge 
transfers possess opposite transfer directions between the spin-up (from the
$K'$ valley to the $K$ one) and -down holes (from the $K$ valley to the $K'$
one). Moreover, as
shown in Fig.~\ref{figyw6} where the steady-state distributions for spin-up
and -down holes in the $K$ valley at $x=1~{\rm \mu{m}}$ are plotted, 
we find that the hole-distributions during the spin diffusion exhibit 
the quasi hot-hole Fermi distribution behaviors, and the 
effective hot-hole temperature of the $k_x>0$
($k_x<0$) states in the distribution [$72~$K ($62~$K)] for spin-down
holes [solid (dashed) curve]
is larger than that [$68~$K ($59~$K)] for spin-up ones [chain (dotted)
  curve], leading to the 
intervalley charge transfer of   
spin-down holes faster than that of spin-up holes. Similar to
Ref.~36 in the time domain, with the weak
hole-phonon scattering but relatively strong hole-hole
Coulomb scattering at low temperature, 
the quasi hot-hole Fermi distributions in the spatial domain are
induced by the spin spatial precession frequency, which
transfers spin-up holes near the corresponding 
Fermi energy into the spin-down states with the same energies, far higher than the
Fermi energy of spin-down holes at large spin polarization. Consequently, as
shown in Fig.~\ref{figyw5}, with
the faster intervalley charge transfer rate of spin-down holes (from the $K$
valley to the $K'$ one), more
holes are accumulated in the $K'$ valley, leading to the valley polarization
built up in the spatial domain.

\subsubsection{Analytical analysis}
\label{anav}
We first focus on the analytical study of the induced valley
polarization in the spatial domain by simplifying
the KSBEs with only the hole-impurity and intervalley hole-phonon scatterings
included. Then the spatial
evolution of the valley polarization $P_{\rm v}=(N_{\rm K'}-N_{\rm K})/{N_{h}}$ is
obtained as (refer to Appendix~\ref{C})     
\begin{equation}
\label{Prate}
\frac{\pi{N_h}}{{m}^2}\frac{{\partial}^2P_{\rm v}}{\partial{x^2}}=\frac{P_{\rm v}}{\tau_p\tau^{+}_v}+\frac{P^{K'}_{s}-P^{K}_{s}}{2\tau_p\tau^{-}_{v}}.
\end{equation} 
Here, $1/\tau^{+(-)}_v$ represents the sum (difference) in the intervalley
charge transfer rates between spin-up and -down holes. 
It can be seen that with the difference in the spin-diffusion lengths in
the two valleys and the difference in the intervalley charge transfer
rates between the spin-up and {-down} holes, the last term $(P^{K'}_{s}-P^{K'}_{s})/(\tau_p\tau^{-}_{v})$
in Eq.~(\ref{Prate}) serves as the source term of the valley
polarization, while the term $P_{\rm v}/(\tau_p\tau^{+}_v)$ in Eq.~(\ref{Prate}) leads to the
relaxation of the valley polarization. 

From Eq.~(\ref{Prate}), the maximum valley polarization $P^{\rm m}_{\rm v}$
along the diffusion direction can 
be approximately obtained as 
\begin{equation}
\label{Pmax}
P^{\rm m}_{\rm v}=\left.\frac{e^{-\beta^{\Downarrow}_{\rm eff}\omega_{\xi}}-e^{-\beta^{\Downarrow}_{\rm eff}\omega_{\xi}}}{e^{-\beta^{\Uparrow}_{\rm eff}\omega_{\xi}}+e^{-\beta^{\Uparrow}_{\rm eff}\omega_{\xi}}}\frac{P^{K'}_{s}-P^{K}_{s}}{2}\right|_{x=x_m}. 
\end{equation}
Here, $\beta^{\Downarrow(\Uparrow)}_{\rm
  eff}=1/(k_BT^{\Downarrow(\Uparrow)}_{\rm eff})$ with $k_B$ the Boltzmann
constant; $\omega_{\xi}$ represents the intervalley phonon energy
($\omega_{K^{\rm L}_6}=17.5~$meV\cite{s_8}). It can been seen from
Eq.~(\ref{Pmax}) that by increasing the difference in the
spin polarizations between the two valleys, $P^{\rm m}_{\rm v}$ can be
enhanced. Moreover, with the
larger difference in the Fermi energies between spin-up and -down holes through 
increasing the injected spin polarization or the hole 
density, the difference in the effective hot-hole
temperatures between spin-up and -down holes becomes larger, leading to the
increase of $P^{\rm m}_{\rm v}$. 

\begin{figure}[htb]
  {\includegraphics[width=8cm]{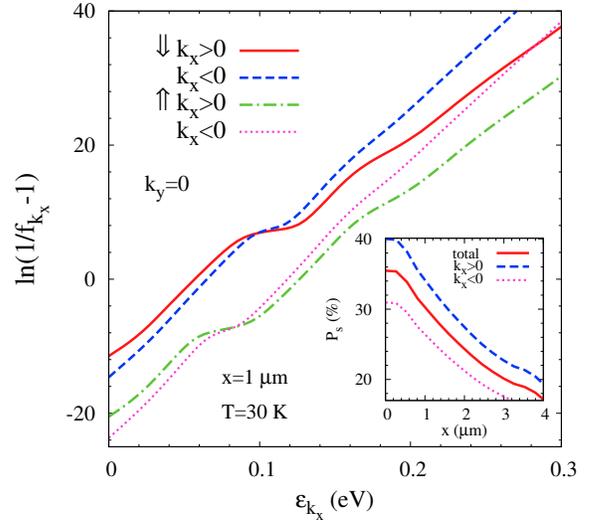}}
\caption{(Color online) 
Hole distribution versus $\varepsilon_{k_x}$ for the states with $k_y=0$
of spin-up and -down holes in the $K$ valley at $x=1~{\rm \mu{m}}$. By
fitting the slope of each curve, the corresponding effective hot-hole
temperature is obtained: $72~$K ($68~$K) for states with $k_x>0$ of spin-up
(-down) holes [solid (chain) curve]; $62~$K 
($59~$K) for states with $k_x<0$ of spin-up (-down)
holes [dashed (dotted) curve]. The inset shows the spin polarizations of
$k_x>0$ states (dashed curve), $k_x<0$
states (dotted curve), and the entire
system (solid curve). $N_h=4\times10^{13}~$cm$^{-2}$ and
$P_s=40~\%$. $E_z=0.03~$V/{\r A}.}      
\label{figyw6}
\end{figure}

The difference in the spin polarizations between
the two valleys is plotted along the diffusion direction in the inset of
Fig.~\ref{figyw7}(a) with all the relevant scatterings
included. Together with the effective hot-hole temperatures
from Fig.~\ref{figyw6}, an estimation of $P^{\rm m}_{\rm v}$ can be obtained
from Eq.~(\ref{Pmax}). However, as seen from Fig.~\ref{figyw6}, it is noted
that for both spin-up and -down holes, the
effective hot-hole temperatures of $k_x>0$ states
are larger than those of $k_x<0$
ones. Specifically, as
shown in Fig.~\ref{figyw5}, at the left edge $x=0$, the 
spin polarization $P^{\mu}_{s,k_x>0}(x=0)$ of the $k_x>0$ states in the distribution is fixed at $P^0_s$ in
our calculation. In the region away from the boundary $(x>0)$,
$P^{\mu}_{s,k_x>0}$ is induced due to the spin injection through the $k_x>0$
states and $P^{\mu}_{s,k_x<0}$ is mainly induced through the scattering from the spin polarized $k_x>0$ states at
the same position $x$. Consequently, in the steady state, with the relatively weak scattering
at low temperature, $P^{\mu}_{s,k_x<0}$ is smaller than
$P^{\mu}_{s,k_x>0}$, as shown in the inset of Fig.~\ref{figyw6}.  
Therefore, with the larger spin polarization and hence the larger difference in the Fermi 
energies between spin-up and -down holes of the $k_x>0$ states in the
distribution, the induced effective hot-hole temperature of the $k_x>0$ states is
larger than that of the $k_x<0$ ones. The anisotropies of the
effective hot-hole temperature and the spin polarization in the distribution
make it complex to obtain an effective hot-hole temperature of the entire
distribution in the spatial domain. Nevertheless, since the quasi hot-hole
Fermi distribution in the spatial domain is very similar to that in the time
domain,\cite{s_8} except for the isotropy of the effective hot-hole temperature in
the distribution in the time domain, 
we approximately take the effective hot-hole temperatures obtained from Ref.~36 in the similar condition, 
and then obtain an estimation of $P^{\rm m}_{\rm v}$ from Eq.~(\ref{Pmax}). Specifically, at
$N_h=4\times10^{13}~$cm$^{-2}$ and 
$T=30~$K, when $P^0_s=30~\%$, with $|P^{K'}_{s}-P^{K}_{s}|_{\rm max}\approx3.5~\%$
[from the inset in Fig.~\ref{figyw7}(a)] and the effective hot-hole
temperatures $T^{\Downarrow}_{\rm eff}=74~$K and $T^{\Uparrow}_{\rm
  eff}=66~$K (obtained from Ref.~36), one has $P^{\rm m}_{\rm v}\approx0.57~\%$ from
Eq.~(\ref{Pmax}). Similarly, for the lager spin injection with $P^0_s=80~\%$ at
$N_h=4\times10^{13}~$cm$^{-2}$ and $T=30~$K, the analytical estimation of
$P^{\rm m}_{\rm v}$ can exceed $1~\%$  
($|P^{K'}_{s}-P^{K'}_{s}|_{\rm max}>5~\%$ with the effective hot-hole
temperatures\cite{s_8} $T^{\Downarrow}_{\rm eff}=150~$K and $T^{\Uparrow}_{\rm
  eff}=112~$K).    

\subsubsection{Numerical results}

\begin{figure}[htb]
  {\includegraphics[width=9cm]{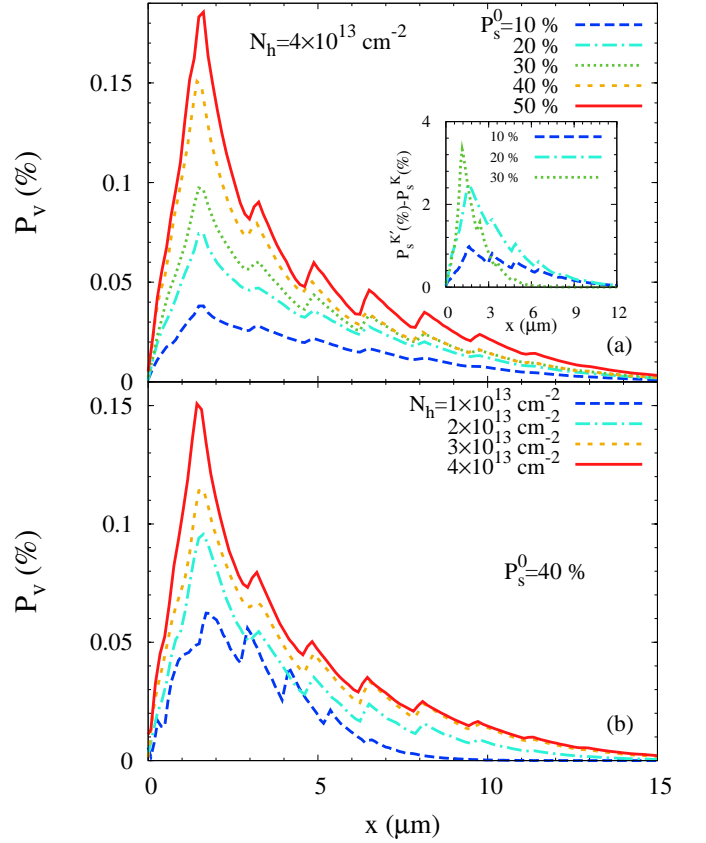}}
\caption{(Color online) The induced valley
polarization $P_{\rm v}$ along the ${\bf {\hat x}}$ direction at different hole
densities and injected spin polarizations when $T=30~$K. The inset in (a) shows the 
difference in the spin polarizations between the two valleys along the diffusion
direction.
}    

\label{figyw7}
\end{figure}

We next discuss the valley polarization by numerically solving the KSBEs with
all the relevant scatterings included. The valley 
polarizations $P_{\rm v}$ along the ${\bf {\hat x}}$ direction
at different hole densities and injected spin polarizations
are plotted in Fig.~\ref{figyw7} when $T=30~$K. To realize the large difference in the
spin-diffusion lengths and hence the spin polarizations in the two
valleys, the electric field in our calculation satisfies
$\eta{E_{z}}=-{\Omega_{\rm HF}}(x=0)$ for given hole density and
injected spin 
polarization $P^0_s$. As seen from the figure, along the ${\bf {\hat x}}$ 
direction, the valley polarization first increases and then decays after
reaching the maximum. This spatial dependence can be understood from Eq.~(\ref{Prate}). 
Near the boundary ($x=0$), the source term
$(P^{K'}_{s}-P^{K'}_{s})/(\tau_p\tau^{-}_{v}){\ne}0$ is more important than
the relaxation one $P_{\rm v}/(\tau_p\tau^{+}_v)$ in
Eq.~(\ref{Prate}), since $P_{\rm v}\approx0$. Therefore, the valley
polarization increases at the first several
micrometers along the ${\bf {\hat x}}$ direction. In the region further away
from the boundary ($x=0$), due to the decay of the spin polarization, the
HF EMF becomes weaker, leading to the smaller difference in the spin-diffusion
lengths in the two valleys. Hence, the source term becomes weaker whereas the
relaxation term is stronger due to the build-up of the valley
polarization. Consequently, the valley polarization starts to decay after
reaching the maximum. Moreover, as shown in Fig.~\ref{figyw7}, the maximum
valley polarization $P^{\rm m}_{\rm v}$ along the ${\bf {\hat x}}$
direction increases with the spin polarization or hole density, qualitatively
consistent with the analytical formula Eq.~(\ref{Pmax}).

However, in contrast to the analytical estimation ($P^{\rm m}_{\rm v}\approx0.57~\%$ at 
$N_h=4\times10^{13}~$cm$^{-2}$ and $T=30~$K for $P^0_s=30~\%$),
it is found that the valley polarization from the numerical
calculation ($P^{\rm m}_{\rm v}\approx0.1~\%$) is much smaller in the same
condition. This is due to the smaller spin polarization than the injected one near the injection
boundary at low temperature. Specifically,  as
mentioned above, in the region away from the boundary $(x>0)$,
$P^{\mu}_{s,k_x>0}$ is induced due to the spin injection through the $k_x>0$
states and $P^{\mu}_{s,k_x<0}$ is induced through the scattering from the spin polarized $k_x>0$ states at
the same position $x$. Consequently, with the relatively weak scattering at low 
temperature, one has
$P^{\mu}_{s,k_x<0}<P^{\mu}_{s,k_x>0}\approx{P^0_s}$ near
the injection boundary ($x\sim0$), 
leading to the spin polarization $P_s$ of the entire distribution smaller than
the injected one $P^0_s$, as shown in the inset of Fig.~\ref{figyw6}. 
For the analytical study of the valley polarization in Sec.~\ref{anav}, the
effective hot-hole temperatures used in the estimation are obtained from
Ref.~36 according to $P^0_s$. Therefore, with the smaller spin polarization
$P_s$ near the injection boundary in the numerical
calculation, the difference in the effective temperatures between the 
spin-up and -down holes is smaller, leading to smaller valley
polarization.

\begin{figure}[htb]
  {\includegraphics[width=9cm]{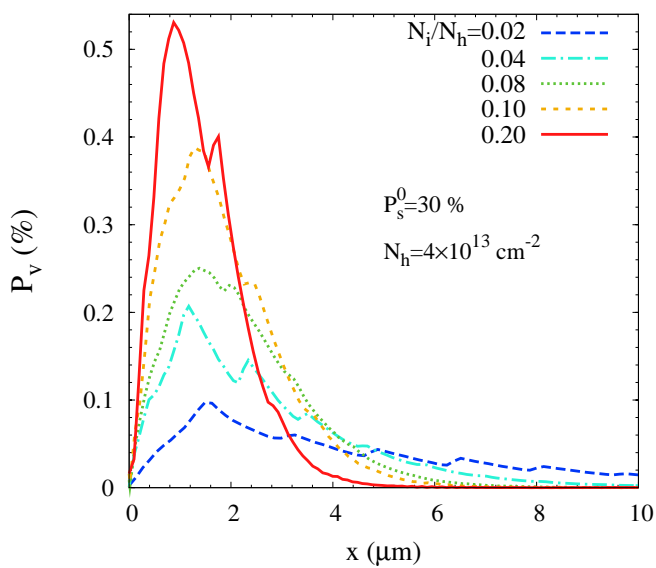}}
  \caption{(Color online) The induced valley
    polarization $P_{\rm v}$ along the ${\bf {\hat x}}$ direction at different
    impurity densities when $N_h=4\times10^{13}~{\rm cm}^{-2}$ and
    $P^0_s=40~\%$. $T=30~$K.}     
  \label{figyw8}
\end{figure}

By increasing impurity density to enhance the scattering strength, $P_s$ near
the injection boundary becomes closer to $P^0_s$, leading to the larger
difference in the effective temperatures between the spin-up and -down holes and
hence the enhanced 
$P^{\rm m}_{\rm v}$. Therefore,  the maximum valley polarization along the ${\bf
  {\hat x}}$ direction increases   
with the impurity density, as shown in Fig.~\ref{figyw8} where the
valley polarizations along the ${\bf {\hat x}}$ direction are plotted at
different impurity densities. This trend is very different from the time
domain, in which the valley polarization always decreases with the increase of the intravalley scattering
strength. Furthermore, it is noted that the increase of the scattering
strength by increasing impurity density also enhances the decay of the
valley polarization after reaching the maximum, which is due to the larger
relaxation [$P_{\rm v}/(\tau_p\tau^{+}_v)$ in Eq.~(\ref{Prate})].

Moreover, as shown in Fig.~\ref{figyw8}, at large impurity density
$N_i/N_h=0.2$, $P^{\rm m}_{\rm v}$ reaches $0.54~\%$ when $P^0_s=40~\%$ with
$T=30~$K and $N_h=4\times10^{13}~{\rm cm}^{-2}$. This
valley polarization from the full numerical calculation is very close to the
simple estimation $P^{\rm m}_{\rm v}\approx0.57~\%$ in the same condition,
and hence confirms the analytical formula Eq.~(\ref{Pmax}). Furthermore, it
is analytically revealed and numerically confirmed above that the larger valley
polarization is expected with the increase of the injected spin polarization or hole
density. Particularly, as mentioned above, from Eq.~(\ref{Pmax}), the estimation
of $P^{\rm m}_{\rm v}$ can exceed $1~\%$ with the injected spin polarization reaching
$80~\%$ when $N_h=4\times10^{13}~$cm$^{-2}$ and $T=30~$K, providing the
possibility for the experimental detection. Unfortunately, a full
numerical computation at very large spin injection ($P^0_s>60~\%$) or
hole density ($N_h>5\times10^{13}~$cm$^{-2}$) needs more grid
points\cite{inho_2} in the momentum and real spaces and goes beyond our
computing power.

\section{SUMMARY}
\label{summary}
In summary, by the KSBE approach with all the relevant scatterings included,
we have investigated the steady-state out-of-plane spin diffusion in $p$-type BL
WSe$_2$ in the presence of the Rashba SOC and HF EMF. The out-of-plane component
of the Rashba SOC serves as the opposite Zeeman-like fields in the two
valleys. Together with the 
identical HF EMFs in the two valleys, the total EMF strengths are different in
the two valleys. The intravalley spin-diffusion processes are shown to play an
important role in the out-of-plane spin diffusion, and due to the
valley-dependent total EMF strength, different intravalley processes in the two
valleys can be obtained.   

Specifically, it is shown that the intravalley spin-diffusion process in each 
valley can be divided into four regimes by tuning the total EMF
strength in the corresponding valley. In different regimes, the spin-diffusion
lengths show different dependencies on the scattering,
total EMF and SOC strengths. At small (large) injected spin polarization and hence the
weak (strong) HF EMFs, the small (large) difference in the total EMF strengths in
the two valleys is obtained, leading to the similar (different)
spin-diffusion lengths in the two valleys. Moreover, we find that the
intervalley hole-phonon scattering can suppress this difference in the
spin-diffusion lengths at large spin injection but becomes
marginal to the spin diffusion at small spin injection. It is further 
revealed that the suppression at large spin injection arises from the 
spin-conserving intervalley charge transfers with the
opposite transfer directions between spin-up and -down
holes by the intervalley hole-phonon
scattering, which tends to suppress the difference in the spin polarizations in
the two valleys. Therefore, with the increase of the intervalley hole-phonon
scattering strength by increasing temperature, the difference in the
spin-diffusion lengths in the two valleys at large spin injection becomes
smaller.    
 
Furthermore, with a fixed single-side large out-of-plane spin injection,
it is found that a steady-state valley polarization along the spin-diffusion
direction is built up at low temperature. Both
analytical and numerical analyses show that it is induced due to the
quasi hot-hole Fermi distributions with different effective hot-hole
temperatures between spin-up and -down holes induced during the spin
diffusion, which leads to the different intervalley charge transfer rates in
the opposite transfer directions, similar to the induced valley polarization
from the spin polarization in the time domain.\cite{s_8} Nevertheless,
different from the maximum valley polarization in the time domain, which
always decreases with increasing the intravalley scattering, the one in the
spatial domain is found to be enhanced by increasing the impurity
density. This unique trend in the spatial domain is because that the
enhancement of the scattering leads to the total spin polarization near the injection
boundary closer to the injected large value, which induces the larger 
difference in the effective hot-hole temperatures between spin-up and -down
holes. The analytical results are
confirmed by the full numerical calculation at 
large impurity density, and it is shown that larger valley
polarization can be reached
by increasing the hole
density or injected spin polarization. Particularly, from the analytical
estimation, the maximum valley polarization along the diffusion direction can
exceed $1~\%$ at the experimental obtainable hole density and with the injected
spin polarization reaching $80~\%$, providing the possibility for the
experimental detection.

\begin{acknowledgments}
This work was supported by the National Natural Science Foundation of
China under Grant No.\ 11334014 and \ 61411136001, the National Basic Research
Program of China under Grant No.\ 2012CB922002, and the Strategic Priority
Research Program of the Chinese Academy of Sciences under Grant No.\
XDB01000000.

\end{acknowledgments}

\begin{appendix}
\section{ANALYTICAL ANALYSIS OF THE SPIN DIFFUSION}
\label{A}
We analytically derive the out-of-plane spin-diffusion length in BL
WSe$_2$ based on the KSBEs for the diffusion along the ${\bf {\hat x}}$
direction. In the steady-state, with only the long-range hole-impurity scattering in the scattering terms, the KSBEs [Eq.~(\ref{KSBEs})]
are written as 
\begin{eqnarray}
\label{KSBEim}
&&k_x\partial_x\rho_{\mu{\bf k}}/m+i\nu{E_z}\big[k_xs_y-k_ys_x,\rho_{\mu{\bf 
  k}}\big]+i\Omega^{\mu}_{\rm T}\big[s_z,\rho_{\mu{\bf
  k}}\big]\nonumber\\
&&\mbox{}+{N_i}\sum_{\bf
  k'}2{\pi}|V_{{\bf k}-{\bf k'}}|^2\delta(\varepsilon_{\bf k}-\varepsilon_{\bf
  k'})(\rho_{\mu{\bf k}}-\rho_{\mu{\bf k'}})=0.
\end{eqnarray}
After the Fourier transformation   
\begin{eqnarray}
\label{Fourier}
 {\rho}^{l}_{\mu{k}}=\frac{1}{2\pi}{\int}^{2\pi}_{0}d\theta_{\bf k}{\rho}_{\mu{\bf
  k}}\exp(-il\theta_{\bf k}),
\end{eqnarray}
Eq.~(\ref{order}) is obtained. In the strong
($l_{\tau}{\ll}l_{\nu},l_{\Omega^\mu_{\rm T}}$) and moderate
($l_{\Omega^\mu_{\rm T}}{\ll}l_{\tau}{\ll}l_{\nu}$) scattering regimes, 
one only needs to keep the lowest two orders ($l=0,1$),\cite{keep,Si,graphene,a1} and has
\begin{eqnarray}
\label{rhoKSBE}
&&l^2_{\tau}\big\{\big(r^2_{\Omega^{\mu}_{\rm
    T}}+1\big)\partial^2_x\rho_{\mu{k}}^{0}-r^2_{\Omega^{\mu}_{\rm
    T}}[s_z,[s_z,\partial^2_x\rho_{\mu{k}}^{0}]] \nonumber\\
&&\mbox{}+2ir_{\Omega^{\mu}_{\rm T}}[s_z,\partial^2_x\rho_{\mu{k}}^{0}]\big\}+2r_{\nu}l^2_{\tau}\big\{ir_{\Omega^{\mu}_{\rm T}}[s_x,\partial_x\rho_{\mu{k}}^{0}]
\nonumber \\
&&\mbox{}+4ir_{\Omega^{\mu}_{\rm T}}[s_z,[is_y,\partial_x\rho_{\mu{k}}^{0}]]+\big(r^2_{\Omega^{\mu}_{\rm
    T}}+2\big)[is_y,\partial_x\rho_{\mu{k}}^{0}]
\nonumber\\
&&\mbox{}-3r^2_{\Omega^{\mu}_{\rm
    T}}[s_z,[s_x,\partial_x\rho_{\mu{k}}^{0}]]\big\}+2r^2_{\Omega^{\mu}_{\rm T}}\big(4r^2_{\nu}+3\big)[s_z,[s_z,\rho_{\mu{k}}^{0}]]\nonumber\\
&&\mbox{}-4r^2_{\nu}\big([s_x,[s_x,\rho_{\mu{k}}^{0}]]-[is_y,[is_y,\rho_{\mu{k}}^{0}]]\big)\nonumber\\  
&&\mbox{}+2ir_{\Omega^{\mu}_{\rm T}}\big(2r^2_{\Omega^{\mu}_{\rm
    T}}-6r^2_{\nu}-1\big)[s_z,\rho_{\mu{k}}^{0}]=0,
\end{eqnarray} 
with $r_{\Omega^{\mu}_{\rm T}}=\Omega^{\mu}_{\rm T}\tau_{k,1}$ and
$r_{\nu}=\nu{kE_z}\tau_{k,1}$.

By defining the spin vector ${\bf S_{\mu}}(x)={\rm Tr}[\rho^0_{\mu{k}}(x){\bm \sigma}]$, the
equation of the out-of-plane spin vector in each valley can be given by
\begin{equation}
\label{sz}
\big(\partial^6_x+3w\partial^4_x/l^2_{\tau}+p\partial^2_x/l^4_{\tau}-q/l^6_{\tau}\big)S_{\mu{z}}=0
\end{equation}
where $w=4r^4_{\Omega^{\mu}_{\rm T}}(1+{\bar r_{\nu}}^2)/3$, $p=16{\bar r_{\nu}}^4+4r^4_{\Omega^{\mu}_{\rm T}}(1+4{\bar r_{\nu}}^2-4{\bar r_{\nu}}^4+r^2_{\Omega^{\mu}_{\rm T}})$ and $q=32r^2_{\nu}(4r^4_{\nu}+r^2_{\Omega^{\mu}_{\rm T}})$
with ${\bar r_{\nu}}^2=r^2_{\nu}/|1+r^2_{\Omega^{\mu}_{\rm T}}|$. 
By solving this equation with the boundary condition
$S_{\mu{z}}(0)=S^0_{\mu{z}}$ and $S_{\mu{z}}(+\infty)=0$, 
the analytical solution of the spin polarization along the ${\bf {\hat x}}$
direction can be obtained as 
\begin{equation}
\label{s}
S_{\mu{z}}(x)=A_o\exp(-x/l^o_s)\cos(x/L_o)+A_s\exp(-x/l^s_s),~~
\end{equation}
with $A_{o(s)}$ being the amplitude for the oscillatory (single-exponential)
decay. The decay length $l^{o(s)}_s$ for the oscillatory (single-exponential)
decay and the oscillation length $L_o$ for the oscillatory decay are
given by  
\begin{eqnarray}
\label{length}
&&l^s_s=l_{\tau}/\sqrt{\Gamma_{\rm r}}, \\
&&l^o_s=\sqrt{2}l_{\tau}/\sqrt{\sqrt{|\Gamma_{+}|^2+|\Gamma_{-}|^2}+\Gamma_{+}},\\
&&L_o=2l^2_{\tau}/|l^o_s\Gamma_{-}|,
\end{eqnarray} 
where
\begin{eqnarray}
&&\Gamma_{\rm r}=-w+\sqrt[3]{b-\sqrt{b^2+d^3}}+\sqrt[3]{b+\sqrt{b^2+d^3}},\\
&&\Gamma_+=-\frac{3}{2}w-\frac{\Gamma_{\rm r}}{2},\\
&&\Gamma_-=\frac{\sqrt{3}}{2}\left(\sqrt[3]{b-\sqrt{b^2+d^3}}-\sqrt[3]{b+\sqrt{b^2+d^3}}\right),~~~~~~~~
\end{eqnarray}
with $b=q/2+w(p-2w^2)/2$ and $d=p/3-w^2$.

It is further found that the above
analytical results can be reduced to simple forms in the four regimes defined in
Sec.~\ref{ana}: I, the large 
total EMF and moderate scattering regime ($l_{\tau}{\ll}l_{\Omega^{\mu}_{\rm
    T}}{\ll}l_{\nu}$); II, the large total EMF and strong scattering regime
($l_{\Omega^{\mu}_{\rm T}}{\ll}l_{\tau}{\ll}l_{\nu}$); III, the crossover regime
($l_{\tau}{\ll}l_{\nu}{\ll}l_{\Omega^{\mu}_{\rm T}}{\ll}2l^2_{\nu}/l_{\tau}$) and
IV, the small total EMF regime
($l_{\tau}{\ll}l_{\nu}{\ll}2l^2_{\nu}/l_{\tau}{\ll}l_{\Omega^{\mu}_{\rm T}}$).
Specifically, for the single-exponential decay 
\begin{eqnarray}
l^s_s\approx
\begin{cases}
(l_{\tau}l_{\nu})/(\sqrt{6}l_{\Omega^{\mu}_T}), &{\rm regime~I}, \cr
l_{\nu}(1+l^2_{\tau}/l^2_{\Omega^{\mu}_T})/\sqrt{2},&{\rm regime~II}, \cr 
l_{\nu}(1-2l^2_{\Omega^{\mu}_T}l^2_{\tau}/l^4_{\nu})/{\sqrt{2}},&{\rm regime~III},  \cr 
l_{\nu}/2,&{\rm regime~IV},
\end{cases}
\end{eqnarray}
for the oscillatory decay
\begin{eqnarray}
l^o_s\approx
\begin{cases}
\sqrt{2}l_{\tau}, &{\rm regime~I}, \cr
l_{\Omega^{\mu}_{T}},&{\rm regime~II}, \cr 
\sqrt{l_{\tau}l_{\Omega^{\mu}_{T}}},&{\rm regime~III},  \cr 
l_{\nu}/(2\sqrt{2\sqrt{2}-1}),&{\rm regime~IV},
\end{cases}
\end{eqnarray}
and the corresponding oscillation length
\begin{eqnarray}
L_o\approx
\begin{cases}
l_{\Omega^\mu_T}/\sqrt{2}, &{\rm regime~I}, \cr
{l_{\tau}},&{\rm regime~II}, \cr 
\sqrt{l_{\tau}l_{\Omega^{\mu}_{T}}},&{\rm regime~III}, \cr 
\sqrt{2\sqrt{2}+1}l_{\nu}/2,&{\rm regime~IV}. 
\end{cases}
\end{eqnarray}

Additionally, one has $A_s\approx{A_o}$ only when $l_s^s{\sim}l^o_s$, and both
the single-exponential decay and oscillatory 
decay are important with the nearly identical decay length. In other cases,
the spin polarization exhibits either one single-exponential decay
or oscillatory decay. In the large
(small) total EMF regime [$l_{\Omega^\mu_T}{\ll}l_{\nu}$
($2l^2_{\nu}/l_{\tau}{\ll}l_{\Omega^\mu_T}$)], the condition for the coexistence 
of single-exponential and oscillatory decays $l_s{\approx}l_o$ is never
satisfied, and hence the steady-state spin polarization is approximated by one
single-exponential (oscillatory) decay in this regime. In the crossover regime
($l_{\nu}{\ll}l_{\Omega^\mu_T}{\ll}2l^2_{\nu}/l_{\tau}$), 
the condition for the coexistence of single-exponential and oscillatory decays
$l^s_s{\sim}l^o_s$ can be satisfied, 
and hence there exists strong competition between the single-exponential and
oscillatory decays in this regime.

\section{ANALYTICAL ANALYSIS OF THE VALLEY POLARIZATION}
\label{C}
We next derive the spatial evolution of the valley polarization in the presence
of the intervalley hole-phonon scattering. The KSBEs [Eq.~(\ref{KSBEs})] with
only the long-range hole-impurity and the intervalley hole-phonon scatterings
are written as
\begin{eqnarray}
\label{KSBEimv}
&&k_x\partial_x\rho_{\mu{\bf k}}/m+i\nu{E_z}\big[k_xs_y-k_ys_x,\rho_{\mu{\bf 
  k}}\big]+i\Omega^{\mu}_{\rm T}\big[s_z,\rho_{\mu{\bf
  k}}\big]\nonumber\\
&&\mbox{}+2{\pi}{N_i}\sum_{\bf
  k'}|V_{{\bf k}{\bf k'}}|^2\delta(\varepsilon_{\bf k}-\varepsilon_{\bf
  k'})(\rho_{\mu{\bf k}}-\rho_{\mu{\bf k'}})+\sum_{\mu'{\bf k'}}|M_{\xi}|^2\nonumber\\
&&\mbox{}\times\big\{\big[(\rho_{\mu{\bf k}}-\rho_{\mu{\bf k'}})n_{\xi}-\rho_{\mu'{\bf k'}}(1-\rho_{\mu{\bf
    k}})\big]\delta(\varepsilon_{\bf k'}-\varepsilon_{\bf k}-\omega_{\xi})\nonumber\\
&&\mbox{}+\big[(\rho_{\mu{\bf k}}-\rho_{\mu{\bf k'}})n_{\xi}+\rho_{\mu{\bf
    k}}(1-\rho_{\mu'{\bf k'}})\big]\delta(\varepsilon_{\bf
  k}-\varepsilon_{\bf k'}-\omega_{\xi})\big\}\nonumber \\
&&\mbox{}\times{2{\pi}}\delta_{\mu',-\mu}=0,
\end{eqnarray}
with $n_{\xi}$ and $|M^{\xi}|$ being the phonon
number and the momentum-independent scattering matrix element\cite{s_8} of the
intervalley phonon $\xi$ mode ($\xi=K^L_6,K^H_6$), respectively.

It is noted that one has $n_{\xi}\approx0$ at low temperature
($k_BT{\ll}\omega_{\xi}$). 
This indicates that the intervalley hole-phonon scattering through
absorbing phonons can be neglected and the one through emitting phonons is important. 
After the Fourier transformation [Eq.~(\ref{Fourier})],
Eq.~(\ref{KSBEimv}) becomes
\begin{eqnarray}
  \label{KF}
&&\frac{k}{2m}\frac{\partial^2}{\partial{x}^2}\big(\rho^{l-1}_{\mu{\bf
    k}}+\rho^{l+1}_{\mu{k}}\big)=-i\Omega^{\mu}_{\rm T}\big[s_z,\rho^{l}_{\mu{k}}\big]-\frac{\rho^l_{\mu{k}}}{\tau_{k,l}}-I^{l}_{\mu{k}}\nonumber\\
&&\mbox{}+\frac{k\nu{E_z}}{2}\big({\big[s_-,{\rho^{l-1}_{\mu{\bf
    k}}}\big]-\big[s_+,{\rho^{l+1}_{\mu{\bf k}}}\big]}\big),
\end{eqnarray}
with
\begin{eqnarray}
\label{I}
&&I^{l}_{\mu{k}}={m|M_{\xi}|^2}\int{\frac{d\epsilon_{k'}}{{2\pi}}}\big[(\rho^l_{\mu{
    k}}(1-\rho^0_{-\mu{k'}})\delta(\varepsilon_{k}-\varepsilon_{k'}-\omega_{\xi})\nonumber\\ 
&&\mbox{} -\rho^0_{-\mu{k'}}(\delta_{l,0}-\rho^l_{\mu{k}})\delta(\varepsilon_{k'}-\varepsilon_{k}-\omega_{\xi})\big].
\end{eqnarray}
Since the intervalley hole-phonon scattering is much weaker than the intravalley
ones, one has ${\rho^l_{\mu{k}}}{\tau^{-1}_{k,l}}{\gg}I^{l}_{\mu{k}}$ in
Eq.~(\ref{KF}) when $l\ne0$ (it is noted that $\tau^{-1}_{k,0}=0$). Therefore,
as mentioned above, in the strong and moderate scattering regimes, one only
needs to keep the lowest two orders $(l=0,1)$, and then obtains\cite{Si,graphene,a1}         
\begin{eqnarray}
\label{rhoKSBEv}
&&l^2_{\tau}\big\{\big(r^2_{\Omega^{\mu}_{\rm
    T}}+1\big)\partial^2_x\rho_{\mu{k}}^{0}-r^2_{\Omega^{\mu}_{\rm
    T}}[s_z,[s_z,\partial^2_x\rho_{\mu{k}}^{0}]] \nonumber\\
&&\mbox{}+2ir_{\Omega^{\mu}_{\rm T}}[s_z,\partial^2_x\rho_{\mu{k}}^{0}]\big\}+2r_{\nu}l^2_{\tau}\big\{ir_{\Omega^{\mu}_{\rm T}}[s_x,\partial_x\rho_{\mu{k}}^{0}]
\nonumber \\
&&\mbox{}+4ir_{\Omega^{\mu}_{\rm T}}[s_z,[is_y,\partial_x\rho_{\mu{k}}^{0}]]+\big(r^2_{\Omega^{\mu}_{\rm
    T}}+2\big)[is_y,\partial_x\rho_{\mu{k}}^{0}]
\nonumber\\
&&\mbox{}-3r^2_{\Omega^{\mu}_{\rm
    T}}[s_z,[s_x,\partial_x\rho_{\mu{k}}^{0}]]\big\}+2r^2_{\Omega^{\mu}_{\rm T}}\big(4r^2_{\nu}+3\big)[s_z,[s_z,\rho_{\mu{k}}^{0}]]\nonumber\\
&&\mbox{}-4r^2_{\nu}\big([s_x,[s_x,\rho_{\mu{k}}^{0}]]-[is_y,[is_y,\rho_{\mu{k}}^{0}]]\big)\nonumber\\  
&&\mbox{}+2ir_{\Omega^{\mu}_{\rm T}}\big(2r^2_{\Omega^{\mu}_{\rm
    T}}-6r^2_{\nu}-1\big)[s_z,\rho_{\mu{k}}^{0}]-6ir_{\Omega^{\mu}_{\rm T}}[s_z,\tau_{k,1}I_{\mu{k}}^{0}]\nonumber\\
&&\mbox{}+6r^2_{\Omega^{\mu}_{\rm T}}[s_z,[s_z,\tau_{k,1}I_{\mu{k}}^{0}]]-2\big(r^2_{\Omega^{\mu}_{\rm T}}+1\big)\tau_{k,1}I_{\mu{k}}^{0}=0.\nonumber\\
\end{eqnarray}
The hole density in each valley $N_{\mu}={\rm Tr}(\rho^0_{\mu{\bf k}})$.
As the hole distribution exhibits a quasi hot-hole Fermi distribution
behavior during the spin diffusion (see Fig.~\ref{figyw6}), we use the
hot-hole Fermi distribution 
$f^{\sigma}_{{\mu}{\bf k}}=1/\{\exp[\beta^{\sigma}_{\rm eff}(\epsilon_{\bf
    k}-\mu_{\mu\sigma})]+1\}$ in the diagonal elements of the density matrices, 
 and then obtain
\begin{equation}
\label{final}
\frac{\pi{N^2_h}}{2{m}^2}\frac{\partial^2}{\partial{x}^2}\big[P_{\rm v}+\frac{{P^{K'}_{s}}^2-{P^{K}_{s}}^2}{4}\big]=\frac{\Delta{N^{\Uparrow}}}{\tau_{k,1}\tau_{v\Uparrow}}-\frac{\Delta{N^{\Downarrow}}}{\tau_{k,1}\tau_{v\Downarrow}}.
\end{equation} 
Here, ${\Delta}N^{\Uparrow(\Downarrow)}=N^{\Uparrow
  (\Downarrow)}_{\rm K'~(K)}-N^{\Uparrow (\Downarrow)}_{\rm K~(K')}$ is the
density difference for 
spin-up (-down) holes between the two valleys; $\tau_{v\Downarrow
  (\Uparrow)}=[e^{\beta(\omega_{\xi}-{\Delta}N^{\Downarrow
    (\Uparrow)}/D_s)}-1]/(2m|M_{\xi}|^2)$ stands for the
intervalley charge transfer time for spin-down (-up) holes with 
$D_s$ being the density of states. 

When all the relevant scatterings are included, $\tau_{k,1}$ in
Eq.~(\ref{final}) is replaced by $\tau_p$. Moreover,
with the suppression on the difference in the spin-diffusion lengths between the two
valleys, we neglect
the second term in the left of Eq.~(\ref{final}), and then Eq.~(\ref{Prate}) is
obtained with
$1/\tau^{\pm}_{v}=1/\tau_{v\Downarrow}\pm1/\tau_{v\Uparrow}$. By assuming 
$\partial^2_xP_{\rm v}=0$ in Eq.~(\ref{Prate}) at $x=x_m$, where
the valley polarization along the diffusion direction reaches the maximum, the
maximum valley polarization [Eq.~(\ref{Pmax})] is obtained.

\end{appendix}

\end{document}